\documentclass[11pt]{article}
\usepackage{amssymb,amsmath,amsfonts}
\usepackage{graphicx}
\usepackage{graphics}
\usepackage{eepic,epsfig}
\usepackage{cite}

\begin{document}

\title{Casimir energy and the possibility of higher dimensional manipulation}
\author{Richard Obousy $^1$, Aram Saharian $^2$ \\
\textit{$^1$ROC, Houston, Texas, 77054, USA}\\
\textit{$^2$Department of Physics, Yerevan State University}\\
\textit{1 Alex Manoogian Street, 0025 Yerevan, Armenia }\\}

\maketitle

\begin{abstract}

It is well known that the Casimir effect is an excellent candidate for the
stabilization of the extra dimensions. It has also been suggested that the Casimir
effect in higher dimensions may be the underlying phenomenon that is responsible
for the dark energy which is currently driving the accelerated expansion of the universe.
In this paper we suggest that, in principle, it may be possible to directly manipulate
the size of an extra dimension locally using Standard Model fields in the next
generation of particle accelerators. This adjustment of the size of the higher
dimension could serve as a technological mechanism to locally adjust the dark
energy density and change the local expansion of spacetime. This idea holds
tantalizing possibilities in the context of exotic spacecraft propulsion.

\end{abstract}

\bigskip

\section{Introduction}

\label{sec:Introd}

Many of the high energy theories of fundamental physics are formulated in
higher dimensional spacetimes. In particular, the idea of extra dimensions
has been extensively used in supergravity and superstring theories. It is
commonly assumed that the extra dimensions are compactified. From an
inflationary point of view, universes with compact spatial dimensions, under
certain conditions, should be considered a rule rather than an exception
\cite{Lind04}. The models of a compact universe with non-trivial topology
may play important roles by providing proper initial conditions for
inflation.

The compactification of spatial dimensions leads to a number of interesting
quantum field theoretic effects, which include instabilities in interacting
field theories, topological mass generation and symmetry breaking. In the
case of non-trivial topology, the boundary conditions imposed on fields give
rise to the modification of the spectrum for vacuum fluctuations and, as a
result, to the Casimir-type contributions in the vacuum expectation values
of physical observables.

The Casimir effect is arguably the most poignant demonstration of the
reality of the quantum vacuum and can be appreciated most simply
in the interaction of a pair of neutral parallel plates. The
presence of the plates modifies the quantum vacuum, and this
modification causes the plates to be pulled toward each other with
a force $F\propto 1/a^4$ where $a$ is the plate separation. The
Casimir effect is a purely quantum phenomenon. In classical
electrodynamics the force between the plates is zero. The ideal
scenario occurs at zero temperature when there are no real photons
(only virtual photons) between the plates; thus, it is the ground
state of the quantum electrodynamic vacuum which causes the
attraction. One of the most important features of the Casimir
effect is that even though it is purely quantum in nature, it
manifests itself macroscopically (Figure 1). \footnote{Illustration
courtesy of Richard Obousy Consulting LLC and AlVin, Antigravit\'{e}} 


\begin{figure}
\begin{center}
\includegraphics[width=11cm, height=7.5cm]{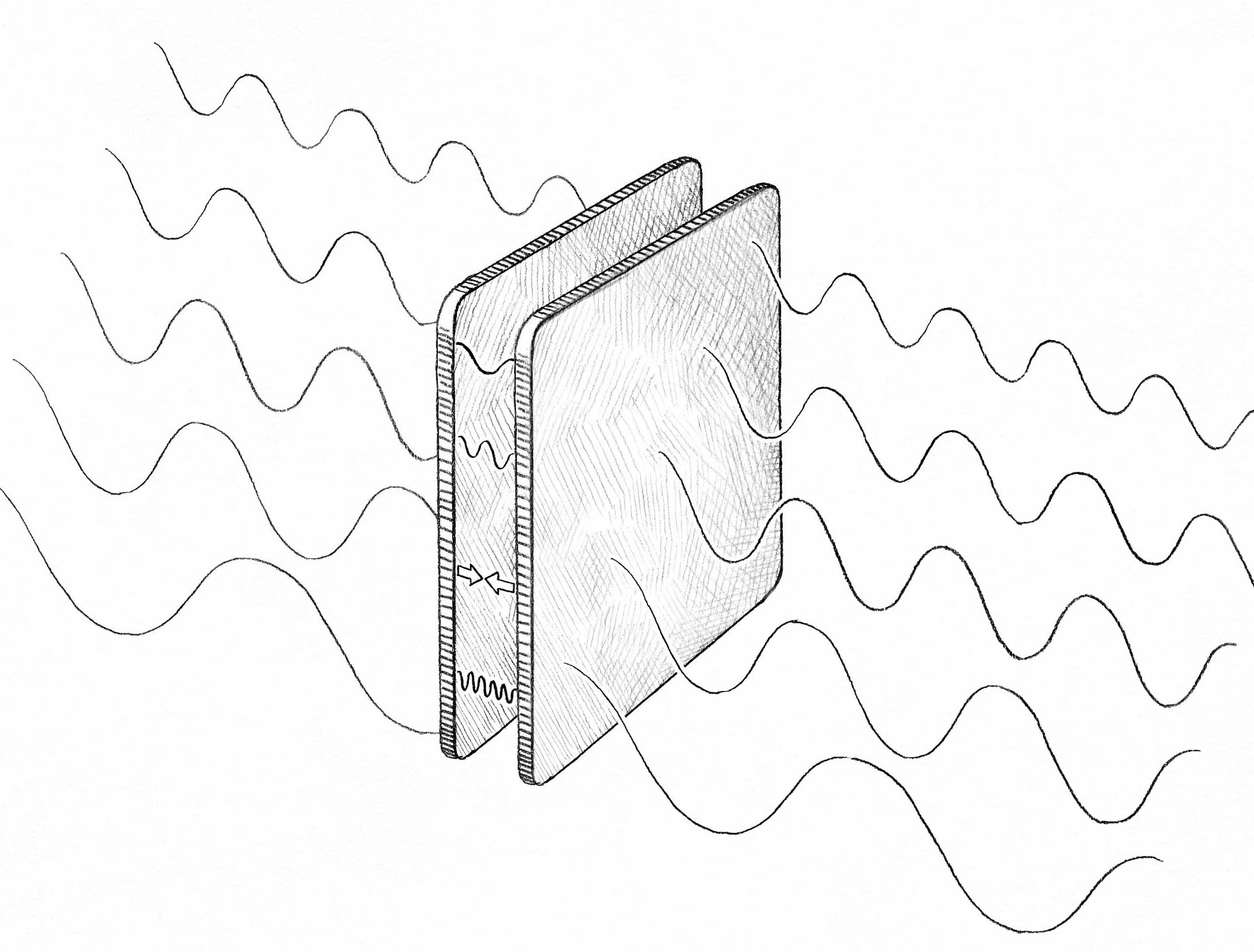}
\caption{Due to the non-trivial boundary conditions imposed on the
quantum vacuum, the plates are pulled toward each other due to a
force that is purely quantum in nature.} \label{fig1}
\end{center}
\end{figure}

Compactifed extra dimensions introduce non-trivial boundary
conditions to the quantum vacuum similar to those in the parallel
plate example, and Casimir-type calculations become important when
calculating the resulting vacuum energy (for the topological
Casimir effect and its role in cosmology see \cite{Most97} and
references therein).

One important question in the context of theories with extra
dimensions is why additional spatial dimensions hold some fixed
size. This is commonly referred to as `modulus stabilization'.
Broadly stated, the question is as follows: if there are
additional spatial dimensions, why do they not perpetually expand,
like our familiar dimensions of space, or alternatively, why do
they not perpetually contract? What mechanism is it that allows
for this higher space to remain compact and stable? Of the handful
of theories that attempt to answer this problem, the Casimir
effect is particularly appealing due to its naturalness.

We find this effect compelling due to the fact that it is a
natural feature intrinsic to the fabric of space itself. There is
much research in the literature that demonstrates that with some
combination of fields, it is possible to generate a stable
minimum of the potential at some extra dimensional radius
\cite{ac}-\cite{Saha05}. In addition, the Casimir effect has been
used as a stabilization mechanism for moduli fields which
parameterize the size and the shape of the extra dimensions both in
models with a smooth distribution of matter in the extra
dimensions and in models with matter located on branes.

The Casimir energy can also serve as a model for dark energy
needed for the explanation of the present accelerated expansion of
the universe (see \cite{Milt03}-\cite{bcp} and references
therein). One interesting feature of these models is the strong
dependence of the dark energy density on the size of the extra
dimension.

One compelling possibility that we would like to introduce in this
paper is that at the energies accessible in the next generation of
particle accelerators, Standard Model (SM) fields might interact
with the graviton Kaluza-Klein (KK) tower which would effect the
local minimum of the potential. This would have the effect of
locally adjusting the radius of the extra dimension during this
interaction. Because the dark energy density is a function of the
size of the extra dimension, one remarkable feature of this idea
is that because the extra dimensions may be (temporarily)
adjusted, so too would the local dark energy density. This
adjustment would mean that, for the duration of the interaction,
the expansion of spacetime would be changed due to
technological intervention. What is particularly appealing about
this predicted phenomenon is its potential application as a future
exotic spacecraft propulsion mechanism.

In the present paper we shortly review the uses of the Casimir
effect in both standard Kaluza-Klein type and braneworld scenarios
for the stabilization of extra dimensions and for the generation
of dark energy. We also explore the energy requirements that would
be needed to temporarily adjust the size of the higher dimension
and hypothesize that, with some imagination, this mechanism could
be used by a sufficiently advanced technology as a means of
spacecraft propulsion.

\section{Kaluza-Klein type models}

To be consistent with the observational data, the extra dimensions in the
standard Kaluza-Klein description are assumed to be microscopic, with a size
much smaller than the scale of four dimensions. A generic prediction of
theories involving extra dimensions is that the gauge and Yukawa couplings
are in general related to the size of extra dimensions. In a cosmological
context, this implies that all couplings depend on the parameters of the
cosmological evolution. In particular, if the corresponding scale factors
are dynamical functions their time dependence induces that for the gauge
coupling constants. However, the strong cosmological constraints on the
variation of gauge couplings coming from the measurements of the quasar
absorbtion lines, the cosmic microwave background, and primordial
nucleosynthesis indicate that the extra dimensions must be not only small
but also static or nearly static. Consequently, the stabilization of extra
dimensions in multidimensional theories near their present day values is a
crucial issue and has been investigated in a number of papers. Various
mechanisms have been considered including fluxes from form-fields, one-loop
quantum effects from compact dimensions, wrapped branes and string corrections.

Let us consider the higher-dimensional action
\begin{equation}
S=\int d^{D}x\sqrt{|\det g_{MN}|}\left\{ -\frac{1}{16\pi G}R\left[ g_{MN}%
\right] +L\right\} ,  \label{action}
\end{equation}%
with the matter Lagrangian $L$ which includes also the cosmological constant
term. We take a spacetime of the form $R\times M_{0}\times \ldots \times
M_{n}$ with the corresponding line element
\begin{equation}
ds^{2}=g_{MN}dx^{M}dx^{N}=g_{\mu \nu }dx^{\mu }dx^{\nu
}+\sum_{i=1}^{n}e^{2\beta _{i}}g_{m_{i}n_{i}}dx^{m_{i}}dx^{n_{i}},
\label{ds2reduc}
\end{equation}%
where $M_{i}$, $i=0,1,\ldots ,n$, are $d_{i}$-dimensional spaces, $g_{\mu
\nu }$, $\beta _{i}$ are functions of the coordinates $x^{\mu }$ in the
subspace $R\times M_{0}$ only, and the metric tensor $g_{m_{i}n_{i}}$ in the
subspace $M_{i}$ is a function of the coordinates $x^{m_{i}}$ in this
subspace only. In order to present the effective action in the subspace $%
R\times M_{0}$ in the standard Einsteinian form we make a conformal
transformation of the $(d_{0}+1)$-dimensional metric:%
\begin{equation}
\tilde{g}_{\mu \nu }=\Omega ^{2}g_{\mu \nu },\;\Omega =\exp \left[
\sum_{j=1}^{n}d_{j}\beta _{j}/(d_{0}-1)\right] .  \label{conftrans}
\end{equation}%
Dropping the total derivatives the action is presented in the form
\begin{equation}
S=\frac{\prod_{j}\mu ^{(j)}}{16\pi G}\int d^{d_{0}+1}x\sqrt{|\det \tilde{g}%
_{\mu \nu }|}\left\{ -R[\tilde{g}_{\mu \nu }]+\sum_{i,j}G_{ij}\tilde{g}^{\mu
\nu }\partial _{\mu }\beta _{i}\partial _{\nu }\beta _{j}-2U\right\} ,
\label{action2}
\end{equation}%
where $\mu ^{(j)}=\int d^{d_{j}}x\sqrt{|\det g_{m_{j}n_{j}}|}$, $%
G_{ij}=d_{i}\delta _{ij}+d_{i}d_{j}/(d_{0}+1)$, and%
\begin{equation}
U=\frac{\Omega ^{-2}}{2\prod_{j}\mu ^{(j)}}\int d^{D-d_{0}-1}x\prod_{j}\sqrt{%
|\det g_{m_{j}n_{j}}|}\left\{ \sum_{i}e^{-2\beta
_{i}}R[g_{m_{i}n_{i}}]-16\pi GL\right\} .  \label{Vpot}
\end{equation}%
In (\ref{Vpot}), $R[g_{m_{i}n_{i}}]$ is the Ricci scalar for the metric $%
g_{m_{i}n_{i}}$.

In the case of standard cosmological metric $\tilde{g}_{\mu \nu }$ with the
scale factor $\tilde{a}_{0}$ and the synchronous time coordinate $\tilde{t}$%
, the field equations for the set of fields $\beta _{i}=\beta _{i}(\tilde{t}%
) $ following from action (\ref{action2}) have the form
\begin{equation}
\sum_{j=1}^{n}G_{ij}\left( \beta _{j}^{\prime \prime }+d_{0}\tilde{\beta}%
_{0}^{\prime }\beta _{j}^{\prime }\right) =-\left( \frac{\partial U}{%
\partial \beta _{i}}\right) _{\tilde{\beta}_{0}},\;i=1,2,\ldots ,n,
\label{coseqnew2}
\end{equation}%
where $\tilde{a}_{0}=\tilde{a}_{0}^{(0)}e^{\tilde{\beta}_{0}}$, $%
a_{j}=a_{j}^{(0)}e^{\beta _{j}}$, the prime means the derivative with
respect to the time coordinate $\tilde{t}$ and the potential $U$ is defined
by the formula
\begin{equation}
U=\frac{1}{2\Omega ^{2}}\left( 16\pi G\rho +\Lambda
_{D}-\sum_{j=1}^{n}\lambda _{j}d_{j}/a_{j}^{2}\right) .  \label{U}
\end{equation}%
In (\ref{U}) $\rho $ is the matter energy density and $\Lambda _{D}$ is the $%
D$-dimensional cosmological constant, $\lambda _{j}=k_{j}(d_{j}-1)$, where $%
k_{j}=-1,0,1$ for the subspace $M_{j}$ with negative, zero, and positive
curvatures, respectively. In the case $\rho =0$ for the extrema of potential
(\ref{U}) one has the relations $\Lambda _{D}=(D-2)\lambda _{i}e^{-2\beta
_{i}}$ and $\partial ^{2}U/\partial \beta _{i}\partial \beta
_{j}=-2G_{ij}\Lambda _{D}/(D-2)$. It follows from here that for $\Lambda
_{D}>0$ the extremum is a maximum of the potential and is realized for
internal spaces with positive curvature. The corresponding effective
cosmological constant is positive. For $\Lambda _{D}<0$ the extremum is a
minimum and is realized for internal spaces with negative curvature. The
effective cosmological constant is negative.

There is a number of mechanisms giving contributions to the potential in
addition to the cosmological constant and curvature terms. An incomplete
list includes fluxes from form-fields, one-loop quantum effects from compact
dimensions (Casimir effect), wrapped branes, string corrections (loop and
classical). In the case of a single extra space for massless fields at zero
temperature the corresponding energy density is of the power-law form
\begin{equation}
\rho =\sigma a_{1}^{-q},  \label{rhocsigma}
\end{equation}%
with a constant $\sigma $ and $q$ being an integer. The values of the
parameter $q$ for various mechanisms are as follows: $q=D$ for the
contribution from the Casimir effect due to massless fields, $q=d_{1}-2p$
for fluxes of $p$-form fields, $q=p-d_{0}-2d_{1}$ for $p$-branes wrapping
the extra dimensions. The corresponding potential has the form (\ref{U})
with
\begin{equation}
U=\frac{1}{2\Omega ^{2}}\left( 16\pi G\sigma a_{1}^{-q}+\Lambda _{D}-\lambda
_{1}d_{1}a_{1}^{-2}\right) .  \label{Uc}
\end{equation}%
In accordance with Eq. (\ref{coseqnew2}), solutions with a static internal
space correspond to the extremum of the potential $U$ and the effective
cosmological constant is related to the value of the potential at the
extremum by the formula $\Lambda _{\mathrm{eff}}=2\Omega ^{2}U$.

\begin{figure}
\begin{center}
\includegraphics[width=8.5cm, height=6.5cm]{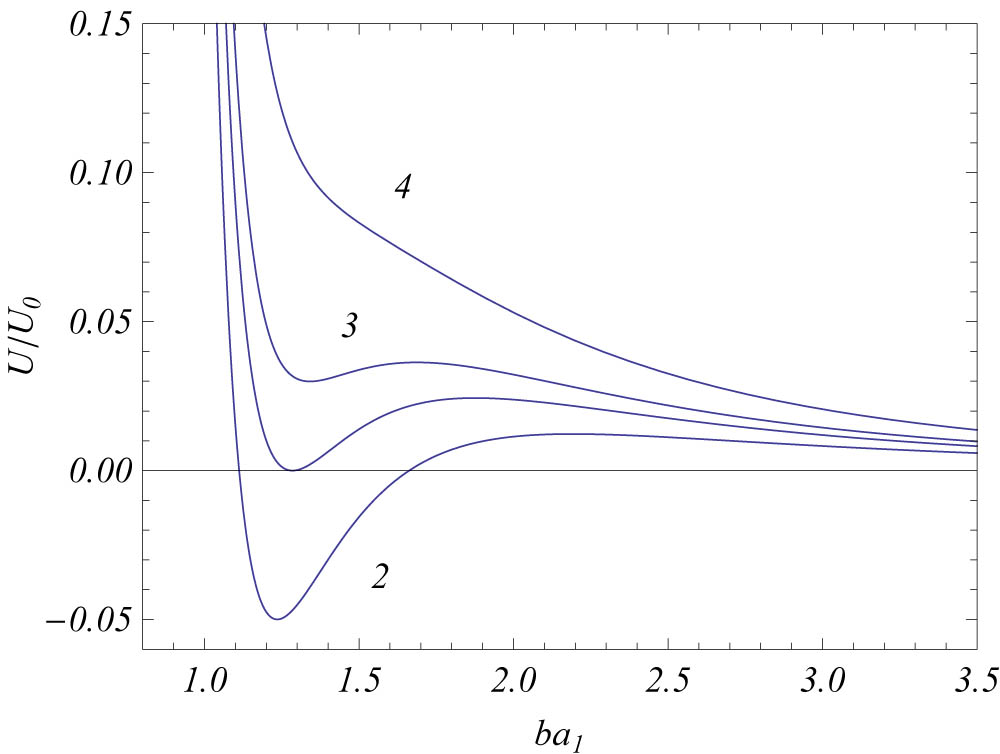}
\caption{The potential $U/U_{0}$ as a function of $ba_{1}$ in the model with
$d_{0}=d_{1}=3$ and with the Casimir effect as a source for $\protect\rho $.
The numbers near the curves correspond to the value of the parameter $%
\Lambda _{D}b^{2}$. For the curve with the zero cosmological constant $%
\Lambda _{D}b^{2}=(30/7)(2/7)^{2/5}$.} \label{fig2}
\end{center}
\end{figure}

Introducing the notations
\begin{equation}
y=ba_{1},\quad b=\left[ \frac{d_{1}(d_{1}-1)}{16\pi G|\sigma |}\right] ^{%
\frac{1}{q-2}},\;U_{0}=\frac{d_{1}(d_{1}-1)}{2}%
b^{2}(ba_{1}^{(0)})^{2d_{1}/(d_{0}-1)},  \label{Uc0}
\end{equation}%
potential (\ref{Uc}) is written in the form
\begin{equation}
U=U_{0}y^{-q-2d_{1}/(d_{0}-1)}\left[ \frac{\Lambda _{D}b^{-2}}{d_{1}(d_{1}-1)%
}y^{q}-k_{1}y^{q-2}+{\mathrm{sign}}(\sigma )\right] .  \label{Ucnew}
\end{equation}%
For the extremum with the zero effective cosmological constant one has%
\begin{equation}
k_{1}y^{q-2}=\frac{q}{2}\,{\mathrm{sign}}(\sigma ),\quad \Lambda
_{D}b^{-2}=k_{1}d_{1}(d_{1}-1)\left( 1-\frac{2}{q}\right) \left( \frac{2}{q}%
\right) ^{\frac{2}{q-2}}.  \label{Lambdab-2}
\end{equation}%
For positive values $q$ and for an internal space with positive (negative)
curvature the extremum exists only when $\sigma >0$ ($\sigma <0$). The
extremum is a minimum for $k_{1}(q-2)>0$ and a maximum for $k_{1}(q-2)<0$.
Potential (\ref{Ucnew}) is a monotonic decreasing positive function for $%
k_{1}=-1,0$, $\Lambda _{D}\geq 0$, $\sigma >0$, has a single maximum with
positive effective cosmological constant for $k_{1}=1,0$, $\Lambda _{D}>0$, $%
\sigma <0$ and $k_{1}=-1$, $\Lambda _{D}\geq 0$, $\sigma <0$. For the case $%
k_{1}=1$, $\Lambda _{D}>0$, $\sigma >0$ and in the $d_{0}=d_{1}=3$ model
where $\rho $ is generated by one-loop quantum effects ($q=D$ in Eq. (\ref%
{rhocsigma})) the potential (\ref{Ucnew}) is plotted in Figure \ref{fig2}
for various values of higher dimensional cosmological constant corresponding
to $\Lambda _{D}b^{-2}=2,(30/7)(2/7)^{2/5},3,4$. In the model with the
second value the effective $(d_{0}+1)$-dimensional cosmological constant
vanishes. The behavior of the potential (\ref{Ucnew}) in the other cases is
obtained from those described above by replacements $(k_{1},\Lambda
_{D},\sigma )\rightarrow (-k_{1},-\Lambda _{D},-\sigma )$ and $U\rightarrow
-U$.


Introducing the $D$-dimensional Planck mass $M_{D}$ in accordance with the
relation $G=M_{D}^{2-D}$, from (\ref{Uc0}) we see that in the model with
one-loop quantum effects the parameter $b$ is of the order of the higher
dimensional Planck mass. As it is seen from the graphs in figure \ref{fig2}
the stabilized value of the size for the internal space is of the order of $%
D $-dimensional Planck length $1/M_{D}$. Note that, as it follows from (\ref%
{action2}), for the effective 4-dimensional Planck mass one has $M_{\mathrm{%
Pl}}^{2}\approx (a_{1}M_{D})^{d_{1}}M_{D}^{2}$ and, hence, in this type of
models the effective and higher dimensional Planck masses have the same
order. Consequently, the size of the internal space is of the order of the
Planck length and is not accessible for the near future accelerators. Note
that our knowledge of the electroweak and strong forces extends with great
precision down to distances of order $10^{-16}$cm. Thus if standard model
fields propagate in extra dimensions, then they must be compactified at a
scale above a few hundred GeV range.

\section{Models with large extra dimensions}

In the previous section we have considered the models, where the extra
dimensions are stabilized by one-loop quantum effects with the combination
of curvature terms and the higher dimensional cosmological constant. We have
taken the Casimir energy density coming from the compact internal space in
the simplest form $\sigma a_{1}^{-D}$. The corresponding effective potential
is monotonic and cannot stabilize the internal space separately. In more
general cases, in particular for massive fields, the dependence of the
Casimir energy on the size of the internal space is more complicated and the
corresponding effective potential can stabilize the extra dimensions
separately.

As a simple model consider the case of a single extra space, $n=1$, being a
circle, $M_{1}=S^{1}$, with a scalar field $\varphi $ as a non-gravitational
source. In the discussion below we will assume that either the external
space $M_{0}$ is non-compact or the external scale factor is much greater
than the internal one. First we discuss the case of the untwisted field with
mass $M$\ satisfying the periodicity condition on $S^{1}$. The vacuum energy
density and the pressures along the uncompactified ($p_{0}$) and
compactified ($p_{1}$) directions are given by the expressions
\begin{eqnarray}
&&\rho =-\frac{2M^{d_{0}+2}}{(2\pi )^{d_{0}/2+2}}\sum_{m=1}^{\infty }\frac{%
K_{d_{0}/2+1}(Mam)}{(Mam)^{d_{0}/2+1}},\quad p_{0}=-\rho ,  \notag \\
&&p_{1}=-\rho -\frac{2M^{d_{0}+2}}{(2\pi )^{d_{0}/2+2}}\sum_{m=1}^{\infty }%
\frac{K_{d_{0}/2+2}(Mam)}{(Mam)^{d_{0}/2}},  \label{s2rhopT02}
\end{eqnarray}%
where $K_{\nu }(x)$ is the modified Bessel function of the second
kind. The corresponding effective potential has the form
\begin{equation}
U=\frac{1}{2}\left( a^{(0)}/a\right) ^{\frac{2}{d_{0}-1}}\left( 16\pi G\rho
+\Lambda _{D}\right) ,  \label{US1}
\end{equation}%
with a constant $a^{(0)}$. The condition for the presence of the static
internal space takes the form
\begin{equation}
d_{0}\rho -(d_{0}-1)p_{1}=\frac{\Lambda _{D}}{8\pi G}.  \label{s2condstat}
\end{equation}%
From formulae (\ref{s2rhopT02}) it follows that the left hand side of this
expression is positive and, hence, the static solutions are present only for
the positive cosmological constant $\Lambda _{D}$. The effective
cosmological constant is equal $\Lambda _{{\mathrm{eff}}}=-8\pi
G(d_{0}-1)(\rho +p_{1})$ and is positive. However, as it can be checked $%
(\partial /\partial a)[d_{0}\rho -(d_{0}-1)p_{1}]<0$ and the solutions with
static internal spaces are unstable.

In the case of a twisted scalar with antiperiodicity condition along the
compactified dimension the corresponding energy density and pressures are
obtained by using the expressions (\ref{s2rhopT02}) with the help of the
formula
\begin{equation}
q_{t}(a)=2q(2a)-q(a),\quad q=\rho ,p_{0},p_{1}.  \label{s2qtw}
\end{equation}%
For the twisted scalar the corresponding energy density is positive. The
solutions with static internal spaces are stable and the effective
cosmological constant is negative: $\Lambda _{{\mathrm{eff}}}<0$.

\begin{figure}
\begin{center}
\includegraphics[width=8.5cm, height=6.5cm]{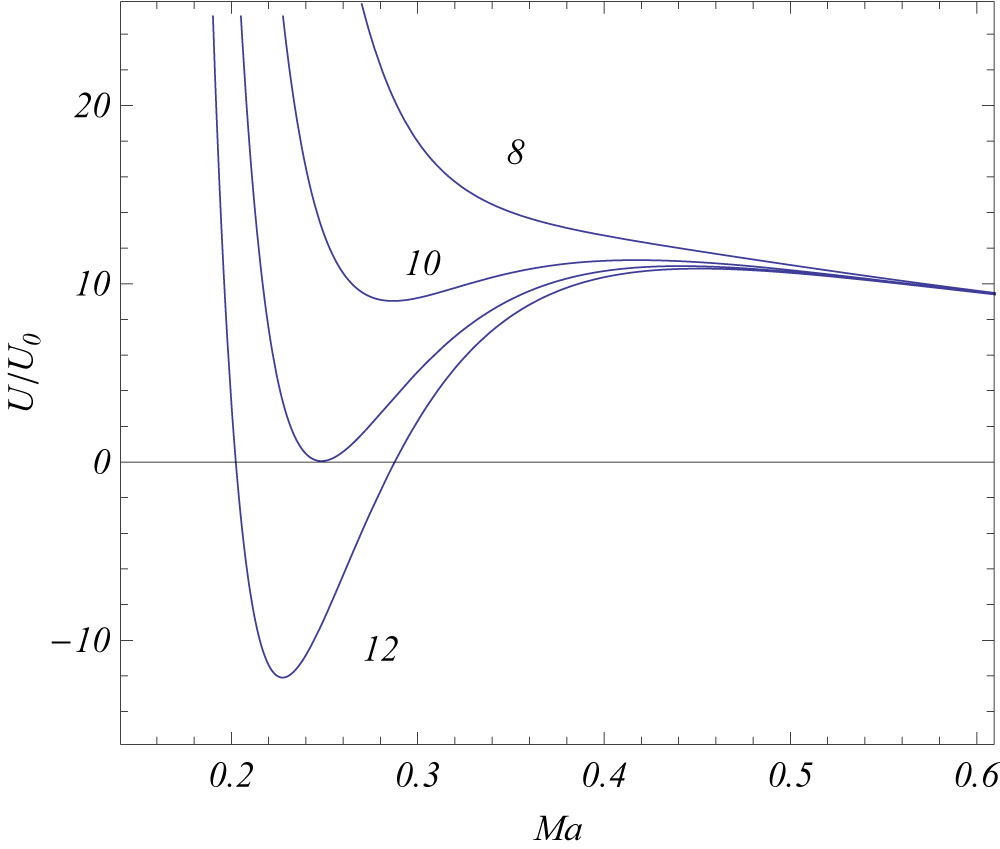}
\caption{The potential $U/U_{0}$ in $d_{0}=3$ as a function of $Ma$ for a
model with single untwisted and two twisted scalars and for $\Lambda _{D}/(8%
\protect\pi GM^{d_{0}+2})=2$. The numbers near the curves correspond to the
values of the ratio $M_{t}/M$. For the graph with zero effective
cosmological constant $M_{t}/M=11.134$.} \label{fig3}
\end{center}
\end{figure}

In Figure \ref{fig3} we have plotted the effective potential $U$ in units of
\begin{equation}
U_{0}=4\pi GM^{d_{0}+2}(Ma^{(0)})^{\frac{2}{d_{0}-1}},  \label{U(0)}
\end{equation}%
as a function on $Ma$ for the model with $d_{0}=3$ consisting a single
untwisted scalar with mass $M$ and two twisted scalars with masses $M_{t}$
for various values of the ratio $M_{t}/M$ (numbers near the curves) and for $%
\Lambda _{D}/(8\pi GM^{d_{0}+2})=2$. For the graph with zero effective
cosmological constant $M_{t}/M\approx 11.134$.

As we see in this type of models the size of the internal space is of the
order $1/M$. The effective cosmological constant is of the order $\Lambda _{{%
\mathrm{eff}}}\sim 8\pi M_{\mathrm{Pl}}^{-2}\times 10^{-2}/a^{4}$. Setting
the value of the corresponding energy density equal to the density of the
dark energy, for the size of the internal space we find $a\sim 10^{-3}%
\mathrm{cm}$. Such large extra dimensions are realized in models where the
standard model fields are localized on a 4-dimensional hypersurface (brane).
Note that the value of the effective cosmological constant can be tuned by
the parameter $M_{t}/M$ and the observed value of the dark energy density
can also be obtained for smaller values for the size of the internal space.
Similar results are obtained in models with more complicated internal
spaces, in \ particular, in ADD models with the number of internal
dimensions $\geqslant 2$.

Note that large extra dimensions accessible to all standard model fields can
also be realized. This type of extra dimensions are referred to as universal
dimensions. The key element in these models is the conservation of momentum
in the Universal Extra Dimensions which leads to the Kaluza-Klein number
conservation. In particular there are no tree-level contributions to the
electrowek observables. The Kaluza-Klein modes may be produced only in
groups of two or more and none of the known bounds on extra dimensions from
single Kaluza-Klein production at colliders applies for universal extra
dimensions. Contribution to precision electroweak observables arise at the
one loop level. In the case of a single extra dimension, recent experimental
constraints allow a compactification scale as low as TeV scale.

\section{Brane models}

Recently it has been suggested that the introduction of compactified extra
spatial dimensions may provide a solution to the hierarchy problem between
the gravitational and electroweak mass scales \cite{Arka98,Rand99}. The main
idea to resolve the large hierarchy is that the small coupling of
four-dimensional gravity is generated by the large physical volume of extra
dimensions. These theories provide a novel setting for discussing
phenomenological and cosmological issues related to extra dimensions. The
model introduced by Randall and Sundrum is particularly attractive. Their
background solution consists of two parallel flat branes, one with positive
tension and another with negative tension embedded in a five-dimensional AdS
bulk \cite{Rand99}. The fifth coordinate is compactified on $S^{1}/Z_{2}$,
and the branes are on the two fixed points. It is assumed that all matter
fields are confined on the branes and only the gravity propagates freely in
the five-dimensional bulk. In this model, the hierarchy problem is solved if
the distance between the branes is about 40 times the AdS radius and we live
on the negative tension brane. More recently, scenarios with additional bulk
fields have been considered.

For the braneworld scenario to be relevant, it is necessary to find a
mechanism for generating a potential to stabilize the distance between the
branes. The braneworld corresponds to a manifold with boundaries and all
fields which propagate in the bulk will give Casimir-type contributions to
the vacuum energy and, as a result, to the vacuum forces acting on the
branes. Casimir forces provide a natural mechanism for stabilizing the
radion field in the Randall-Sundrum model, as required for a complete
solution of the hierarchy problem. In addition, the Casimir energy gives a
contribution to both the brane and bulk cosmological constants and, hence,
has to be taken into account in the self-consistent formulation of the
braneworld dynamics.

In the $(D+1)$-dimensional generalization of the Randall-Sundrum spacetime
the background spacetime is described by the line-element%
\begin{equation}
ds^{2}=e^{-2k|y|}\eta _{\mu \nu }dx^{\mu }dx^{\nu }-dy^{2},  \label{ds2AdS}
\end{equation}%
where $\eta _{\mu \nu }$ is the metric tensor for the $D$-dimensional
Minkowski spacetime the AdS curvature radius is given by $1/k$. The fifth
dimension $y$ is compactified on an orbifold, $S^{1}/Z_{2}$ of length $a$,
with $-a<y<a$. The orbifold fixed points at $y=0$ and $y=a$ are the
locations of two $D$-branes. Consider a scalar field $\varphi (x)$\ with
curvature coupling parameter $\zeta $ obeying boundary conditions $\left(
\tilde{A}_{y}+\partial _{y}\right) \varphi (x)=0$, $y=0,a$, on the branes.
For a scalar field with brane mass terms $c_{0}$ and $c_{a}$ on the left and
right branes respectively, the coefficients in the boundary conditions are
defined by the relation $2\tilde{A}_{j}=-n^{(j)}c_{j}-4D\zeta k$ with $%
n^{(0)}=1$, $n^{(b)}=-1$ (see, for instance, \cite{Gher00,Flac01,Saha05}).
The corresponding Casimir energy is given by the expression \cite%
{Gold00,Flac01,Garr01,Saha05}
\begin{equation}
E(a)=\alpha +\beta e^{-Dka}+\frac{(4\pi )^{-D/2}k^{D}}{\Gamma \left(
D/2\right) }\int_{0}^{\infty }du\,u^{D-1}\ln \left\vert 1-\frac{\bar{I}_{\nu
}^{(a)}(u)\bar{K}_{\nu }^{(b)}(ue^{ka})}{\bar{K}_{\nu }^{(a)}(u)\bar{I}_{\nu
}^{(b)}(ue^{ka})}\right\vert ,  \label{ECasbr}
\end{equation}%
where $I_{\nu }(u)$ and $K_{\nu }(u)$ are the modified Bessel
functions,
\begin{equation}
\nu =\sqrt{(D/2)^{2}-D(D+1)\zeta +m^{2}/k^{2}},  \label{nu}
\end{equation}%
and the barred notation for a given function $F(x)$ is defined by $\bar{F}%
^{(j)}(x)=(\tilde{A}_{j}/k+D/2)F(x)+xF^{\prime }(x)$. The first and second
terms on the right of Eq. (\ref{ECasbr}) are the energies for a single brane
located at $y=0$ and $y=a$ respectively when the second brane is absent. The
coefficients $\alpha $ and $\beta $ cannot be determined from the low-energy
effective theory and can be fixed by imposing suitable renormalization
conditions which relate them to observables. The last term in (\ref{ECasbr})
is free of renormalization umbiguities and can be termed as an interaction
part. The corresponding vacuum forces acting on the branes can be either
repulsive or attractive in dependence of the coefficients in the boundary
conditions. In particular, there is a region in the parameter space where
these forces are attractive at large distances between the branes and
repulsive at small distances with an equilibrium state at some intermediate
distance $a=a_{0}$. In the original Randall-Sundrum braneworld with $D=4$,
to account for the observed hierarchy between the gravitational and
electroweak scales we need $ka_{0}\approx 37$.

In addition to the stabilization of the distance between the branes, the
quantum effects from bulk fields can also provide a mechanism for the
generation of the cosmological constant on the visible brane. On manifolds
with boundaries in addition to volume part, the vacuum energy contains a
contribution located on the boundary. For a scalar field the surface energy
density $\varepsilon _{\mathrm{v}}^{{\mathrm{(surf)}}}$\ on the visible
brane $y=a$ is presented as the sum \cite{Saha04}
\begin{equation}
\varepsilon _{\mathrm{v}}^{{\mathrm{(surf)}}}=\varepsilon _{1\mathrm{v}}^{{%
\mathrm{(surf)}}}+\Delta \varepsilon _{\mathrm{v}}^{{\mathrm{(surf)}}},
\label{emt2pl2}
\end{equation}%
where $\varepsilon _{1\mathrm{v}}^{{\mathrm{(surf)}}}$ is the surface energy
density on the visible brane when the hidden brane is absent, and the part
\begin{equation}
\Delta \varepsilon _{\mathrm{v}}^{{\mathrm{(surf)}}}=k^{D}\frac{\left(
4\zeta -1\right) \tilde{A}_{j}-2\zeta }{(4\pi )^{D/2}\Gamma \left(
D/2\right) }\int_{0}^{\infty }du\,\frac{u^{D-1}\bar{I}_{\nu
}^{(a)}(ue^{-ka})/\bar{I}_{\nu }^{(b)}(u)}{\bar{K}_{\nu }^{(a)}(ue^{-ka})%
\bar{I}_{\nu }^{(b)}(u)-\bar{I}_{\nu }^{(a)}(ue^{-ka})\bar{K}_{\nu }^{(b)}(u)%
},  \label{Delteps}
\end{equation}%
is induced by the presence of the hidden brane. The part $\varepsilon _{1%
\mathrm{v}}^{{\mathrm{(surf)}}}$ contains finite renormalization terms which
are not computable within the framework of the model under consideration and
their values should be fixed by additional renormalization conditions. The
effective cosmological constant generated by the hidden brane is determined
by the relation
\begin{equation}
\Lambda _{\mathrm{v}}=8\pi M_{\mathrm{v}}^{2-D}\Delta \varepsilon _{\mathrm{v%
}}^{{\mathrm{(surf)}}},  \label{effCC}
\end{equation}%
where $M_{\mathrm{v}}$ is the $D$-dimensional effective Planck mass scale
for an observer on the visible brane. For large interbrane distances the
quantity (\ref{effCC}) is suppressed compared with the corresponding Planck
scale quantity in the brane universe by the factor $\exp [-k(2\nu +D)a]$,
assuming that the AdS inverse radius and the fundamental Planck mass are of
the same order. In the original Randall-Sundrum model with $D=4$, for a
scalar field with the mass $|m|^{2}\lesssim k^{2}$, and interbrane distances
solving the hierarchy problem, the value of the induced cosmological
constant on the visible brane by the order of magnitude is in agreement with
the value of the dark energy density suggested by current cosmological
observations without an additional fine tuning of the parameters.

\section{Exotic Propulsion}

The idea of manipulating spacetime in some ingenious fashion to
facilitate a novel form of spacecraft propulsion has been well
explored in the literature \cite{oc1,oc2,mty,Ford:1994bj,
Ford:1996er, Pfenning:1997wh, Pfenning:1997rg, Everett:1997hb,
Everett:1995nn, Visser:1998ua, VanDenBroeck:1999sn, Lobo:2004wq,
Hiscock:1997ya, GonzalezDiaz:1999db, Puthoff:1996my,
Natario:2004zr, Lobo:2004an, Bennett:1995gp, GonzalezDiaz:2007zzb,
alc, DavisMillis:2009}. If we are to realistically entertain the
notion of interstellar exploration in timeframes of a human
life-span, a profound shift in the traditional approach to
spacecraft propulsion is clearly necessary. It is well known that
the universe imposes dramatic relativistic limitations on all
bodies moving through spacetime, and that all matter is restricted
to motion at sublight velocities ($< 3\times10^8  \rm{m/s}$, the
speed of light, or c) and that as matter approaches the speed of
light, its mass asymptotically approaches infinity. This mass
increase ensures that an infinite amount of energy would be
necessary to travel at the speed of light, and thus, this speed is
impossible to reach and represents an absolute speed limit to all
matter travelling through spacetime.

Even if an engine were designed that could propel a spacecraft to
an appreciable fraction of light-speed, travel to even the closest
stars would take many decades in the frame of reference of an
observer located on Earth. Although these lengthy transit times
would not make interstellar exploration impossible, they would
certainly reduce the enthusiasm of governments or private
individuals funding these missions. After all, a mission whose
success is, perhaps, a century away would be difficult to justify.
In recent years, however, two loop-holes to Einstein's ultimate
speed limit are known to exist: the Einstein-Rosen bridge  and the
warp-drive. Fundamentally, both ideas involve the manipulation of
spacetime itself in some exotic way that allows for superluminal
travel.

The warp drive, which is the main feature of study for this
section, involves local manipulation of the fabric of space in the
immediate vicinity of a spacecraft. The basic idea is to create an
asymmetric bubble of space that is contracting in front of the
spacecraft while expanding behind it. \footnote{Recent progress by
Jos\'{e} Nat\'{a}rio has demonstrated that, with a slightly more
complicated metric, one can dispense of the expansion.} Spacetime
is believed to have stretched at many times the speed of light in
the first second of its existence during the inflationary period.
In many ways, the idea presented in this paper is an artifical and
local re-creation of those initial conditions.

Using this form of locomotion, the spacecraft remains stationary
inside this `warp bubble,' and it is the movement of space itself
that facilitates the relative motion of the spacecraft.  The most
attractive feature of the warp drive is that the theory of
relativity places no known restrictions on the motion of space
itself, thus allowing for a convenient circumvention of the speed
of light barrier.


\begin{figure}
\begin{center}
\includegraphics[width=8cm, height=6.8cm]{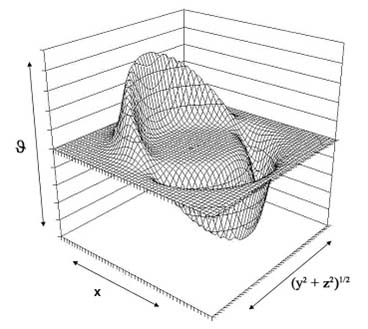}
\caption{The `Top hat metric'. A bubble of asymmetric spacetime
curvature surrounds a spacecraft which would sit in the center of
the bubble. The space immediately surrounding the
spacecraft would be expanding/contracting behind/in front of the
craft. In this image the ship would `move' from left to right.} \label{fig4}
\end{center}
\end{figure}

By associating dark energy with the Casimir energy due to the KK
modes of vacuum fluctuations in higher dimensions, especially in
the context of M-theory derived or inspired models, it is possible
to form a relationship between $\Lambda$ and the radius of the
compact extra dimension.
\begin{equation}
\rho=\Lambda\propto 1/a^D.
\end{equation}
An easier way of developing the relationship between the energy
density and the expansion of space is to discuss quantities in
terms of Hubble's constant H, which describes the rate of
expansion of space per unit distance of space.
\begin{equation}
H\propto\sqrt{\Lambda} ,
\end{equation}
or in terms of the radius of the extra dimension we have
\begin{equation}
H\propto1/a^{D/2}.
\end{equation}

\begin{figure}
\begin{center}
\includegraphics[width=11.5cm, height=7.5cm]{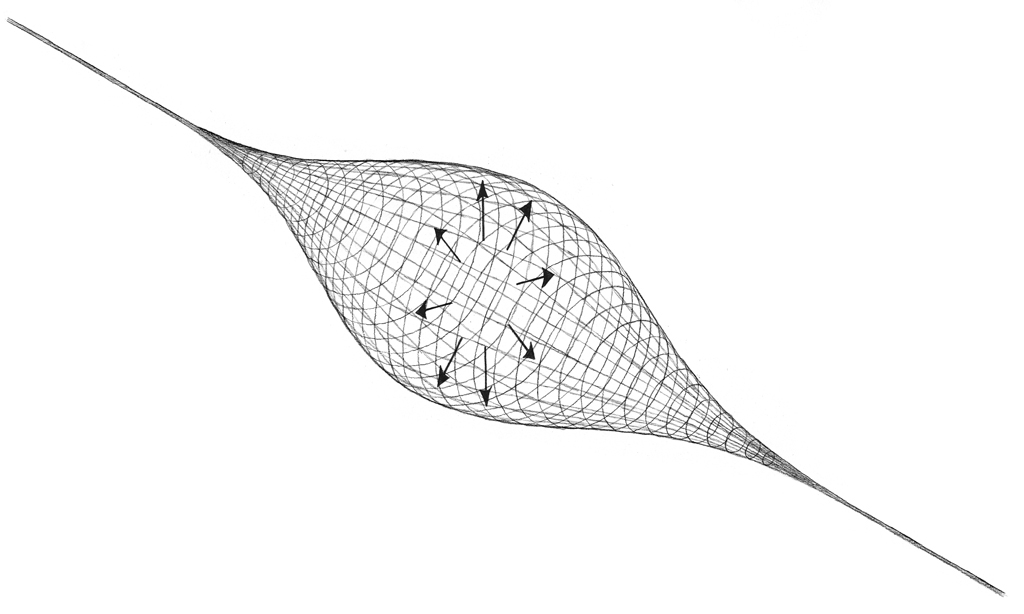}
\caption{An extra dimension that is artifically stimulated to
contract (or expand), due to interaction with the SM fields, might
provide a way to locally adjust the dark energy density.} \label{fig5}
\end{center}
\end{figure}

This result indicates that within this model, the expansion of
spacetime is a function of the size of the higher dimension. One
fascinating question to ask at this point is: could it be possible
to effect the radius of a higher dimension through some advanced
technology? If this were, indeed, possible then this technology
would provide a remarkable mechanism to locally adjust the dark
energy density, and thus the local expansion rate of spacetime
(Figure \ref{fig5}). \footnote{Illustration
courtesy of Richard Obousy Consulting LLC and AlVin, Antigravit\'{e}}

In principal, this represents an interesting way to artificially
generate the type of spacetime featured in Figure \ref{fig4}. A
spacecraft with the ability to create such a bubble would always
move inside its own local light-cone. This ship could utilize the
expansion of spacetime behind the ship to move away from some
object at any desired speed, or equivalently, to contract the
space-time in front of the ship to approach any object. The
possibility that the size of the compact geometry might, indeed,
vary depending on the location in four dimensional spacetime has
been explored in the context of string theory \cite{Giddings2005},
but never from the perspective of propulsion technology.


To explore the types of energies that may be required to generate
this manipulation of a higher dimension, we consider that fact
that in models with large extra dimensions, the interaction of the
graviton KK tower with the Standard Model fields are suppressed
by the higher dimensional Planck scale, and the corresponding
couplings are inverse TeV in strength. This can be seen more
clearly when we consider the expansion of the metric tensor in
models with large extra dimensions computed within linearized
gravity models:
\begin{equation}
g_{MN}=\eta_{MN}+h_{MN}/M_D^{D/2-1},
\end{equation}
where $\eta_{MN}$ corresponds to flat (Minkowski) spacetime and
$h_{MN}$ corresponds to the bulk graviton fluctuations. The
graviton interaction term in the action is expressed by:
\begin{equation}
S_{\rm{int}}=1/M_D^{D/2-1}\int d^D x h_{MN}T^{MN}
\end{equation}
with $T^{MN}$ being the higher dimensional energy-momentum tensor.
The interaction of the graviton KK states with the SM fields are
obtained by integrating the action over the extra coordinates.
Because all these states are coupled with the universal strength $
1/M_{\rm{Pl}}$ this leads to the compelling possibility of the
control of the size of the extra dimensions by processes at
energies that will be accessible via the particle accelerators of
the near future. Although the coupling is extremely small, the
effective coupling is enhanced by the large number of KK states.

Referring to \ref{fig2}, additional energy in the form of matter
or radiation with the TeV energy scale can alter the shape of the
effective potential. In particular, the extrema determining the
size of the extra dimensions are modified with the change of the
Casimir energy density and hence, the dark energy density, in the
models under consideration.


\begin{figure}
\begin{center}
\includegraphics[width=7.52cm, height=4.76cm]{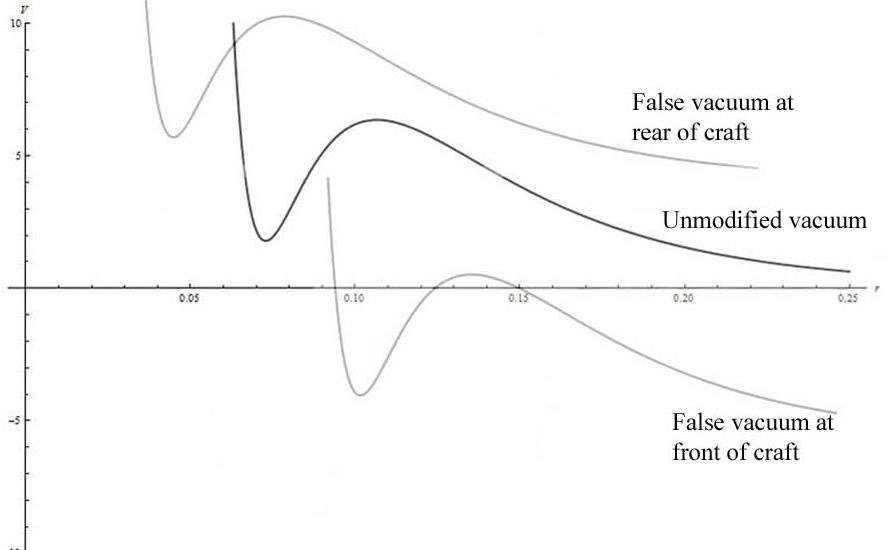}
\caption{By surrounding a spacecraft with false
minima an artificial inflation of the local spacetime might be
achieved.} \label{fig6}
\end{center}
\end{figure}

The key to creating a warp drive in this model is to create a
false vacuum minimum, i.e., to modify the vacuum spectrum and
inject some field which creates a deSitter minimum at the rear of
the craft and an anti-deSitter minimum at the front of the craft.
What this requires is a technology that would allow us to
artificially manipulate the field content illustrated in
\ref{fig2}, shifting the location of the minimum. In this basic
representation, the spacecraft would sit in a stable region of
space corresponding to the natural minimum of the extra dimension
(approximately flat space). At the front and rear of the craft,
regions of false minima would be artificially created via the
adjustment of the extra dimension (Figure \ref{fig6}). These modified
regions would correspond to increased and negative dark energy
densities, thus creating the warp bubble previously discussed.

\section{Extra Dimensions in General Relativity}

In the context of GR a similar phenomenology is produced for the
case of anisotropic cosmological models, in which it is the
$\textit{contraction}$ of the extra dimension that has the effect
of expanding another \cite{Levin:1994}. For example, consider a
`toy' universe with one additional spatial dimension with the
following metric
\begin{equation}
ds^2=dt^2-a^2(t)d\vec{x}^2-b^2(t)dy^2  \ .
\end{equation}
In this toy universe we will assume spacetime is empty, that there
is no cosmological constant, and that all spatial dimensions are
locally flat,
\begin{equation}
T_{\mu \nu}=\Lambda g_{\mu \nu}=0 \ .
\end{equation}
The action of the Einstein theory of gravity generalized to five dimensions will be
\begin{equation}
S^{(5)}=\int d^4x dy \sqrt{-g^{(5)}}\left(\frac{M_5^2 }{16\pi }R^{(5)}\right) \ .
\end{equation}
Solving the vacuum Einstein equations
\begin{equation}
G_{\mu\nu}=0 \ ,
\end{equation}
we obtain for the $G_{11}$ component
\begin{equation}
G_{11}=\frac{3\dot{a}(b\dot{a}+a\dot{b})}{a^2b} \ .
\end{equation}
Rewriting $\dot{a}/a=H_a$ and $\dot{b}/b=H_b$ where $H_a$ and
$H_b$ corresponds to the Hubble constant in three space and the
Hubble constant in the extra dimension respectively, we find that
solving for $G_{11}=0$ yields
\begin{equation}
H_a=-H_b .
\end{equation}
This remarkable result indicates that in a vacuum, the shear of a
contracting dimension is able to inflate the remaining dimensions.
In other words, the expansion of the 3-volume is associated with
the contraction of the one-volume.

Even in the limit of flat spacetime with zero cosmological
constant, general relativity shows that the physics of the
$\textit{compactified}$ space affects the expansion rate of the
non-compact space. The main difference to note here is that the
quantum field theoretic result demonstrates that a
$\textit{fixed}$ compactification radius can also result in
expansion of the three-volume due to the Casimir effect, whereas
the GR approach suggests that a $\textit{changing}$
compactification radius results in expansion, as is shown in (34).
What is particularly interesting about this result is that it
demonstrates that the expansion and contraction of the non-compact
space seems to be intimately related to the extra dimension in
both QFT and GR calculations.

\section{Discussion}

As we have seen, the stabilization of compact extra dimensions and the
acceleration of the other 4-dimensional part of the spacetime can be
simultaneously described by using the Casimir energy-momentum tensor as a
source in the Einstein equations. The acceleration in the 3-dimensional
subspace occurs naturally when the extra dimensions are stabilized. In this
case the Casimir energy density $\rho _{C}$ is a constant with the equation
of state $p_{C}=-\rho _{C}$, where $p_{C}$ is the Casimir pressure in the
visible universe, and an effective cosmological constant is induced. Note
that the current observational bounds for the equation of state parameter
are $-1.4<$ $p_{\mathrm{DE}}/\rho _{\mathrm{DE}}<-0.85$ and the value -1 of
this parameter is among the best fits of the recent observational data. For
the Casimir energy density to be the dark energy driving accelerated
expansion the size of the internal space should be of the order $10^{-3}%
\mathrm{cm}$. Such large extra dimensions are realized in braneworld
scenario. The value of the effective cosmological constant can be tuned by
choosing the masses of the fields and the observed value of the dark energy
density can also be obtained for smaller values for the size of the internal
space.

An important feature of both models with a smooth distribution of
matter in the extra dimensions and with branes is the dependence
of the dark energy density on the size of the extra dimensions. In
models with large extra dimensions the interaction of the graviton
Kaluza-Klein tower with the standard model fields are suppressed
by the higher dimensional Planck scale and the corresponding
couplings are inverse TeV strength. This leads to the interesting
possibility for the control of the size of extra dimensions by
processes at energies accessible at the near future particle
accelerators. This is seen from the form of the effective
potential for the scale factor of the internal subspace given
above. Additional energy density in the form of matter or
radiation with the TeV energy scale can alter the shape of the
effective potential. In particular, the extrema determining the
size of the extra dimensions are changed with the change of the
Casimir energy density and, hence, the dark energy density in the
models under consideration.

In the ADD scenario the Standard model
fields are confined to a 4D brane and a
new gravity scale $M_{D}=G^{1/(2-D)}\gtrsim \mathrm{TeV}$ is introduced in $%
4+d_{1}$ dimensions. The gravity on the brane appears as a tower of
Kaluza-Klein states with universal coupling to the Standard Model fields.
Though this coupling is small ($\sim 1/M_{\mathrm{Pl}}$), a relatively large
cross section is obtained from the large number of Kaluza-Klein states.
Kaluza-Klein graviton emission from SN1987A puts a bound $M_{D}\geqslant 30\,%
\mathrm{TeV}$ on the fundamental energy scale for ADD type models
in the case $d_{1}=2$. For the models with $d_{1}=3$ the
constraint is less restrictive, $M_{D}\geqslant
\mathrm{few\,\,TeVs}$. Hence, the latter case is still viable for
solving the hierarchy problem and accessible to being tested at
the LHC. The ATLAS experiment which will start taking data at the
LHC will be able to probe ADD type extra dimensions up to
$M_{D}\approx $ $8$ TeV. In the Randall-Sundrum scenario the
hierarchy is explained by an exponential warp factor in
$\mathrm{AdS}_{5}$ bulk geometry. Electroweak precision tests put
a severe lower bound on the lowest Kaluza-Klein mass. The masses
of the order of $1\,\mathrm{TeV}$ can be accommodated. In
universal extra dimension scenarios the current Tevatron results
constrain the mass of the compactification scale to
$M_{\mathrm{C}}>400$ GeV. The ATLAS experiment will be sensitive
to $M_{\mathrm{C}}\approx 3$ TeV. These estimates show that if the
nature is described by one of the models under consideration, the
extra dimensions can be probed and their size can be controlled at
the energies accessible at LHC. In models with the Casimir energy
in the role of the dark energy this provides a possibility for the
control of the dark energy local density through the change in the
size of the extra dimensions.

Note that the effective potential (\ref{Vpot}) contains a factor $\Omega
^{-2}\sim V_{{\mathrm{int}}}^{-2/(d_{0}-1)}$, with $V_{{\mathrm{int}}}$
being the volume of the internal space, and, hence, vanishes in the large
volume limit for the extra dimensions if the energy density grows not
sufficiently fast. This leads to the instability of a dS minimum with
respect to the decompactification of extra dimensions. This feature is
characteristic for other stabilization mechanisms as well including those
based on fluxes from form-fields and wrapped branes. The stability
properties of the metastable dS minimum depend on the height of the local
maximum separating this minimum from the minimum which corresponds to the
infinite size of the extra dimensions. Now we conclude that in models with
large extra dimensions, as in the case of the parameters for the minimum,
the height and the location of the maximum can be controlled by the
processes with the energy density of the TeV scale.

Another important point which should be touched is the following.
Though the size of the internal space is stabilized by the
one-loop effective potential, in general, there are fluctuations
around the corresponding minimum with both classical and quantum
sources which shift the size for the internal space from the fixed
value leading to the time variations of this size. This gives time
variations in the equation of state parameter for the
corresponding dark energy density. However, it should be taken
into account that the variations of the size for the internal
space around the minimum induces variation of fundamental
constants (in particular, gauge couplings) in the effective
four-dimensional theory. These variations are strongly constrained
by the observational data. As a consequence, the variation of the
equation of state parameter for the corresponding dark energy
density around the mean value $-1$ are small. It is interesting
that in the models under consideration the constancy of the dark
energy density is related to the constancy of effective physical
constants. Note that models can be constructed where the volume of
the extra dimensions is stabilized at early times, thus
guaranteeing standard four dimensional gravity, while the role of
quintessence is played by the moduli fields controlling the shape
of the extra space.


\begin{thebibliography}{99}
\bibitem{Lind04} A. Linde, {\it Creation of a compact topologically nontrivial
inflationary universe}, JCAP \textbf{10}, 004 (2004).

\bibitem{Most97} V.M. Mostepanenko and N.N. Trunov, \textit{The Casimir
effect and its applications} (Clarendon, Oxford, 1997); M. Bordag,
U. Mohideen and V.M. Mostepanenko, {\it New developments in the
Casimir effect}, Phys. Rept. \textbf{353}, 1 (2001); K.A. Milton,
\textit{The Casimir effect: physical manifestation of zero-point
energy} (World Scientific, Singapore, 2002); E. Elizalde, S.D.
Odintsov, A. Romeo, A.A. Bytsenko and S. Zerbini, \textit{Zeta
regularization techniques with applications} (World Scientific,
Singapore, 1994); M.J. Duff, B.E.W. Nilsson and C.N. Pope, {\it
Kaluza-Klein Supergravity}, Phys. Rept. \textbf{130}, 1 (1986);
A.A. Bytsenko, G. Cognola, L. Vanzo, and S. Zerbini, {\it Quantum
Fields and Extended Objects in Space-Times with Constant Curvature
Spatial Section}, Phys. Rept. \textbf{266}, 1 (1996).


\bibitem{ac}T.\ Appelquist and A.\ Chodos, {\it The Quantum Dynamics Of Kaluza-Klein Theories},
Phys. Rev. D \textbf{28}, 772 (1983).

\bibitem{ac1}T.\ Appelquist and A.\ Chodos, {\it Quantum Effects In Kaluza-Klein Theories},
Phys. Rev. Lett. \textbf{50}, 141 (1983).

\bibitem{rr}M.\ Rubin and B.\ Roth, {\it Fermions and Stability in Five Dimensional Kaluza Klein Theory},
PLB \textbf{127}, 127 (1983).

\bibitem{I}T.\ Inami, {\it Quantum Effects In Generalized Kaluza-Klein Theories},
Phys. Rev. B \textbf{133}, 180 (1983).

\bibitem{Blau84} S.K. Blau, E.I. Guendelman, A. Taormina, L.C.R.
Wijewardhana, {\it On the Stability of Toroidally Compact
Kaluza-Klein Theories}, Phys. Lett. B \textbf{144}, 30--36 (1984).

\bibitem{Cand84} P. Candelas, S. Weinberg, {\it Calculation of Gauge Couplings and Compact
Circumferences From Self-Consistent Dimensional Reduction}, Nucl.
Phys. B \textbf{237}, 397--441 (1984).

\bibitem{Gilb84} G. Gilbert, B. McClain, {\it Fermions and Stability in Quantum Kaluza-Klein
Theories}, Nucl. Phys. B \textbf{244}, 173--185 (1984).

\bibitem{Kikk85} K. Kikkawa, T. Kubota, S. Sawada, M. Yamasaki, {\it Spontaneous
Compactification in Generalized Candelas-Weinberg Models}, Nucl.
Phys. B \textbf{260}, 429--455 (1985).

\bibitem{Acce86} F.S. Accetta, M. Gleiser, R. Holman, E.W. Kolb,
{\it Towards Stable Compactifications}, Nucl. Phys. B
\textbf{276}, 501--516 (1986).

\bibitem{Maed87} K. Maeda, {\it Stability and Attractor in a
Higher-Dimensional Cosmology}, Class. Quantum Grav. \textbf{3},
233--247 (1986).

\bibitem{Gunt97} U. G\"{u}nther, A. Zhuk, {\it Gravitational Excitons
from Extra Dimensions}, Phys. Rev. D \textbf{56}, 6391--6402
(1997); U. G\"{u}nther, A. Zhuk, {\it Stabilization of Internal
Spaces in Multidimensional Cosmology}, Phys. Rev. D \textbf{61},
124001 (2000).

\bibitem{nos} S.\ Nojiri, S.\ Odintsov and S.\ Zerbini {\it Bulk versus
boundary (gravitational Casimir) effects in quantum creation of
inflationary brane world universe}, Class. Quant. Grav.
\textbf{17}, 4855 (2000); S. Nojiri and S.D. Odintsov, {\it Brane
world inflation induced by quantum effects}, Phys. Lett. B
\textbf{484}, 119 (2000).

\bibitem{Gold00} W.\ Goldberger and I.\ Rothstein, {\it Quantum
stabilization of compactified $AdS_5$}, Phys. Lett. B \textbf{491},
339 (2000).

\bibitem{Garr01} J.\ Garriga, O.\ Pujolas, and T.\ Tanaka, {\it Radion
effective potential in the brane-world}, Nucl. Phys. B
\textbf{605}, 192 (2001).

\bibitem{Pont01} E. Pont\'{o}n, E. Poppitz, {\it Casimir Energy and Radius
Stabilization in Five and Six Dimensional Orbifolds}, JHEP
\textbf{06}, 019 (2001).

\bibitem{Flac01} A. Flachi and D.J. Toms, {\it Quantized bulk scalar
fields in the Randall-Sundrum brane-model}, Nucl. Phys. B
\textbf{610}, 144 (2001).


\bibitem{m} S.\ Mukohyama, {\it Quantum effects, brane tension and
large hierarchy in the brane world}, Phys. Rev. D \textbf{63},
044008 (2001).

\bibitem{fmt}A.\ Flachi, I.\ Moss, and D.\ Toms, {\it Quantized bulk fermions
in the Randall-Sundrum brane model}, Phys. Rev. D \textbf{64},
105029 (2001).

\bibitem{hkp}R.\ Hofmann, P.\ Kanti, and M.\ Pospelov {\it (De)stabilization
of an extra dimension due to a Casimir force}, Phys. Rev. D
\textbf{63}, 124020 (2001).

\bibitem{kt}H.\ Kudoh and T.\ Tanaka, {\it Second order perturbations in the
Randall-Sundrum infinite brane world model}, Phys. Rev. D
\textbf{64}, 084022, 2001.

\bibitem{nos1}S.\ Nojiri, S.\ Odintsov, and S.\ Sugamoto, {\it Stabilization and
radion in de Sitter brane world}, Mod. Phys. Lett. A \textbf{17},
1269 (2002).

\bibitem{enoo} E.\ Elizalde, S.\ Nojiri, S.D.\ Odintsov, and S.\ Ogushi, {\it Casimir
effect in de Sitter and anti-de Sitter brane worlds}, Phys. Rev. D \textbf{67}, 063515, (2003).

\bibitem{gpt1} J.\ Garriga, O.\ Pujolas, and T.\ Tanaka, {\it Moduli effective
action in warped brane world compactifications}, Nucl. Phys. B \textbf{655},127 (2003)

\bibitem{Saha04a} A.A. Saharian and M.R. Setare, {\it The Casimir Effect on
Background of Conformally Flat Brane-World Geometries}, Phys.
Lett. B \textbf{552}, 119 (2003).

\bibitem{Saha05} A.A. Saharian, {\it Wightman function and Casimir
densities on AdS bulk with application to the Randall-Sundrum
braneworld}, Nucl. Phys. B \textbf{712}, 196 (2005).


\bibitem{Milt03} K.A. Milton, {\it Dark Energy as Evidence for Extra Dimensions},
Grav. Cosmol. \textbf{9}, 66 (2003); E. Elizalde, {\it Uses of
Zeta Regularization in QFT with Boundary Conditions: a
Cosmo-Topological Casimir Effect}, J. Phys. A \textbf{39}, 6299
(2006).

\bibitem{gl}B.\ Greene and J.\ Levin, {\it Dark Energy and Stabilization of
Extra Dimensions}, JHEP \textbf{0711}, 96 (2007).

\bibitem{bcp}P.\ Burikham, A.\ Chatrabhuti, and P.\ Patcharamaneepakorn,
{\it  Dark energy and moduli stabilization of extra dimensions in
M(1+3) T2 spacetime}, JHEP \textbf{07}, 013 (2008).

\bibitem{Arka98} N. Arkani-Hamed, S. Dimopoulos, and G. Dvali,
{\it The Hierarchy Problem and New Dimensions at a Millimeter},
Phys. Lett. B \textbf{429}, 263 (1998); N. Arkani-Hamed, S.
Dimopoulos, and G. Dvali, {\it Phenomenology, Astrophysics and
Cosmology of Theories with Sub-Millimeter Dimensions and TeV Scale
Quantum Gravity}, Phys. Rev. D \textbf{59}, 086004 (1999); I.
Antoniadis, N. Arkani-Hamed, S. Dimopoulos, and G. Dvali, {\it New
Dimensions at a Millimeter to a Fermi and Superstrings at a TeV},
Phys. Lett. B \textbf{436}, 257 (1998).

\bibitem{Rand99} L. Randall and R. Sundrum, {\it A Large Mass Hierarchy From
a Small Extra Dimension}, Phys. Rev. Lett. \textbf{83}, 3370
(1999); L. Randall and R. Sundrum, {\it An Alternative to
Compactification}, Phys. Rev. Lett. \textbf{83}, 4690 (1999).

\bibitem{Gher00} T. Gherghetta and A. Pomarol, {\it Bulk Fields and
Supersymmetry in a Slice of AdS}, Nucl. Phys. B \textbf{586},
141 (2000).

\bibitem{Saha04} A.A. Saharian, {\it Surface Casimir Densities and
Induced Cosmological cConstant on Parallel Branes in AdS}, Phys.
Rev. D \textbf{70}, 064026 (2004).



\bibitem{oc1}
R.~Obousy and G.~Cleaver, {\it Warp Drive: A New Approach}, JBIS
\textbf{61}, 149 (2008).

\bibitem{oc2}
R.~Obousy and G.~Cleaver, {\it Putting the `Warp' into Warp
Drive}, Spaceflight \textbf{50}, 149 (2008).

\bibitem{mty}
M.~Morris, K.~Thorne, and U.~Yurtsever, {\it Wormholes, Time
Machines, and the Weak Energy Condition}, Phys. Rev. Lett.
\textbf{61}, 1466 (1988).

\bibitem{Ford:1994bj}
L.~H. Ford and T.~A. Roman, {\it Averaged Energy Conditions and
Quantum Inequalities}, Phys. Rev. D \textbf{51}, 4277 (1995).

\bibitem{Ford:1996er}
L.~H. Ford and T.~A. Roman, {\it Restrictions on Negative Energy
Density in Flat Spacetime}, Phys. Rev. D \textbf{55}, 2082 (1997).

\bibitem{Pfenning:1997wh}
M.~J. Pfenning and L.~H. Ford, {\it The Unphysical Nature of `Warp
Drive'}, Class. Quant. Grav. \textbf{14}, 1743 (1997).

\bibitem{Pfenning:1997rg}
M.~J. Pfenning and L.~H. Ford, {\it Scalar Field Quantum
Inequalities in Static Spacetimes}, Phys. Rev. D \textbf{57}, 3489
(1998).

\bibitem{Everett:1997hb}
A.~E. Everett and T.~A. Roman, {\it A Superluminal Subway: The
Krasnikov Tube}, Phys. Rev. D \textbf{56}, 2100 (1997).

\bibitem{Everett:1995nn}
A.~E. Everett, {\it Warp Drive and Causality}, Phys. Rev. D
\textbf{53}, 7365 (1996).

\bibitem{Visser:1998ua}
M. Visser, B. Bassett, and S. Liberati, {it Superluminal
Censorship}, Nucl. Phys. Proc. Suppl. \textbf{88}, 267 (2000).

\bibitem{VanDenBroeck:1999sn}
C. Van Den~Broeck, {\it A `Warp Drive' with Reasonable Total
Energy Requirements}, Class. Quant. Grav. \textbf{16}, 3973
(1999).

\bibitem{Lobo:2004wq}
F.~S.~N. Lobo and M. Visser, {\it Fundamental Limitations on 'Warp
Drive' spacetimes}, Class. Quant. Grav. \textbf{21}, 5871 (2004).

\bibitem{Hiscock:1997ya}
W.~A. Hiscock, {\it Quantum Effects in the Alcubierre Warp Drive
Spacetime}, Class. Quant. Grav. \textbf{14}, L183 (1997).

\bibitem{GonzalezDiaz:1999db}
P.~F. Gonzalez-Diaz, {\it On the Warp Drive Space-Time}, Phys.
Rev. D \textbf{62}, 044005 (2000).

\bibitem{Puthoff:1996my}
H.~E. Puthoff, {\it {SETI}, the Velocity-of-Light Limitation, and
the Alcubierre Warp Drive: An Integrating Overview}, Phys. Essays,
\textbf{9}, 156 (1996).

\bibitem{Natario:2004zr}
J. Natario, {\it Newtonian Limits of Warp Drive Spacetimes}, Gen.
Rel. Grav. \textbf{38}, 475 (2006).

\bibitem{Lobo:2004an}
F.~S.~N. Lobo and M. Visser, {\it Linearized Warp Drive and the
Energy Conditions}, gr-qc/0412065 (2004).

\bibitem{Bennett:1995gp}
G.~L. Bennett, R.~L. Forward, and R.~H. Frisbee, {\it Report on
the NASA/JPL Workshop on Advanced Quantum/Relativity Theory
Propulsion}, AIAA Paper, \textbf{1995N2599}, 1 (1995).

\bibitem{GonzalezDiaz:2007zzb} P.~F. Gonzalez-Diaz,
{\it Superluminal Warp Drive and Dark Energy}, Phys. Lett. B
\textbf{657}, 15 (2007).

\bibitem{alc} M.~Alcubierre, {\it The Warp Drive: Hyperfast Travel Within General Relativity},
Class. Quant. Grav. \textbf{11}, L73 (1994).

\bibitem{DavisMillis:2009} E. Davis, M. Millis, {\it Frontiers of Propulsion Science},
AIAA Press, Reston, VA, pp. 473--509, 2009.


\bibitem{Giddings2005}
S.~Giddings, {\it Gravity and Strings}, Lecture given at 32nd SLAC
Summer Institute on Particle Physics, vol. SSI, 2004.

\bibitem{Levin:1994}J. J. Levin, {\it Inflation from Extra Dimensions},
Phys. Lett. B \textbf{343}, 69 (1995).

\end{thebibliography}
\end{document}